\def\ltsim{ \,{}^<_\sim\, }

\def\ts{\thinspace}

\def\simle{\thinspace\hbox{\raise.3ex\hbox{$<$} \llap{$_\sim$~}}\negthinspace}
\def\simge{\thinspace\hbox{\raise.3ex\hbox{$>$} \llap{$_\sim$~}}\negthinspace}

\newcommand{\caiii}{\mbox{$\lambda_{8498}$}}
\newcommand{\caii}{\mbox{$\lambda_{8542}$}}
\newcommand{\cai}{\mbox{$\lambda_{8662}$}}

\newcommand{\ca}{\ion{Ca}{2}}

\newcommand{\ews}{\mbox{$\Sigma Ca$}}

\newcommand{\fe}{\mbox{[Fe/H]}}
\newcommand{\cah}{\mbox{[Ca/H]}}
\newcommand{\cafe}{\mbox{[Ca/Fe]}}

\newcommand{\vhb}{\mbox{$V_{HB}$}}

\newcommand{\redew}{\mbox{$W^{\prime}$}}
\newcommand{\vhbmv}{\mbox{$V_{HB} - V$}}
\newcommand{\mAA}{\mbox{\rm \AA}}
\newcommand{\logg}{\mbox{\rm log(g)}}
\newcommand{\me}{\mbox{$m.e.1$}}
\newcommand{\teff}{\mbox{$\rm T_{eff}$}}
\newcommand{\lee}{\mbox{$(B-R)/(B+V+R)$}}
\newcommand{\qzinn}{\mbox{$Q_{39}$}}
\newcommand{\ie}{\mbox{\em i.e.}}

\documentstyle[11pt,aaspp4]{article}

\lefthead{Rutledge et al.}
\righthead{Globular Cluster Ca II Metallicities}

\begin{document}

\title{Galactic Globular Cluster Metallicity Scale From the \ca\ Triplet \\
       II. Rankings, Comparisons and Puzzles}

\author{Glen A. Rutledge,  James E. Hesser, and Peter B. Stetson}
\affil{National~Research~Council~of~Canada,
       Herzberg~Institute~of~Astrophysics,
       Dominion~Astrophysical~Observatory, 
        5071 W. Saanich Rd., 
        RR5, Victoria, BC V8X~4M6, Canada 
	\\ Electronic mail: firstname.lastname@hia.nrc.ca}

\begin{abstract}
We compare our compilation of the \redew\ calcium index  for 71
Galactic globular clusters to the widely used \markcite{zinn84}Zinn and
West (1984 ApJS, 55, 45) \fe\ scale and to Carretta and Gratton's (1997
A\&AS, 121, 95) scale from high-dispersion spectra analyzed with Kurucz
(1992, private communication) model atmospheres.  We find our calcium 
ranking to be tightly
correlated with each comparison set, in a non-linear  and a linear
fashion, respectively. By combining our calcium index information with
the Zinn and West ranking, we are able to rank the globular clusters in
our sample with a typical precision of $\pm 0.05$ dex for $\fe_{ZW84}
\lesssim -0.5$;  for clusters more metal rich than this, the ranking is
less precise. The significant differences between these metallicity
scales raise important questions about our understanding of Galactic
formation and chemical enrichment processes. Furthermore, in spite of
the apparent improvement in metallicity ranking for the Galactic
globular clusters that results from our addition of information from
the \ca\ triplet lines to the potpourri of other metallicity
indicators, caution -- perhaps considerable -- may be advisable when using
\redew\ as a surrogate for metallicity, especially for systems where
ranges in age and metallicity are likely.

\end{abstract}

\pagebreak[0]
\section{Introduction} \label{intro}

In Paper I of this series (\markcite{rutledge97}Rutledge et al. 1997),
we presented new observations of the \ca\ triplet lines at
\caiii, \caii, and \cai\ in the spectra of 976 stars lying near the red
giant branches (RGBs) of 52 Galactic globular clusters. The aim of this
work is to establish a relative ranking of the clusters according to
their calcium abundances, and to apply this ranking to some of the
astrophysical problems associated with understanding the early stages
of Galactic formation. 

In \S \ref{indicator}, we summarize how the \ca\ triplet is used to
produce a calcium index, \redew, for a given cluster, and how this
index is related to the metallicity, \fe, of the cluster. A catalog of
\redew\ values for 71 clusters is created in \S \ref{comp} from a
compilation of data from Paper I and from other studies. Zinn and West
(1984, ZW84) produced a relative ranking of the globular clusters using
a compilation of almost all the ranking techniques available at that
time, to which we make detailed comparisons in \S \ref{relative}.
Considering the past debate over the effect of HB 
morphology on \markcite{zinn80b} Zinn's (1980b) photometric
\qzinn\ index (\markcite{smith84}Smith 1984, \markcite{frogel83}Frogel
et al. 1983, \markcite{zinn84}ZW84), which plays a significant role in the
$\fe_{ZW84}$ scale,  we have re-investigated this
issue in \S \ref{q39} using our precise \redew\ index and more recent
HB types from \markcite{lee94} Lee et al. (1994).  We
then make a detailed comparison of our calcium ranking with the
\fe\ ranking of ZW84, which prove to be highly correlated, but in a \em
significantly non-linear \rm way.  By combining the information in both
systems, we are able to provide a metallicity ranking of improved
precision (\S \ref{zw84}).  In \S \ref{absolute}, we compare our
calcium ranking to \fe\ values from high dispersion spectroscopy
studies.  We transform our calcium index to \fe\ on the
\markcite{carretta97}Carretta and Gratton (1997, CG97) scale, which is
highly correlated with our \redew\ index in  an \em essentially
linear \rm ~fashion.  Finally, in \S \ref{discuss}, we briefly discuss
the significance of these disparate  metallicity scales and the use of
the \ca\ triplet lines as a metallicity indicator when studying
Galactic formation physics.

\section{\ca\ Triplet as a Metallicity Indicator} \label{indicator}

The \ca\ triplet lines at \caiii, \caii, and \cai\ are used to probe
various astrophysical phenomena because they are among the strongest
features in the near infrared spectra of most late-type stars and
stellar systems, and thus their equivalent widths can be measured
reasonably accurately for faint objects with moderate resolution
spectrographs.  Most of the early integrated light work took advantage
of the sensitivity in the triplet's summed equivalent width (denoted as
\ews) to \logg\ as a means to discriminate between dwarf and giant
populations (\markcite{spinrad69}\markcite{spinrad71}Spinrad and Taylor
1969, 1971, \markcite{anderson74} Anderson 1974).  Initially, it was
thought that \ews\ was relatively insensitive to metallicity
(\markcite{cohen79}Cohen 1978, 1979, \markcite{jones84}Jones et al.
1984), but \markcite{alloin89} Alloin and Bica's (1989) re-analysis of
\markcite{jones84} Jones et al.'s data found that metallicity does have
a significant effect on \ews.  \markcite{diaz89} Diaz et al. (1989)
found, from a sample of 106 bright stars with spectral classes F6-M0,
that 97\% of the variance in \ews\ resulted from a linear combination
of \logg\ and \fe\, with little dependence on \teff.  They suggested
that \fe\ only plays a significant role when $\fe\ \lesssim -0.3$~dex;
otherwise, \logg\ is the dominant factor.  The sensitivity of \ews\ to
\fe\ was not fully appreciated until it was measured for Galactic
globular clusters and dwarf spheroidals.

Inspired by the success of the \ca\  K line to rank globular
clusters according to \fe\ (\markcite{zinn84}ZW84),
\markcite{armzinn88}Armandroff and Zinn (1988, AZ88) measured
\ews\ in the integrated light of 27 globular clusters and found very
good correlations with other metallicity-sensitive indices. 
Since there is an intrinsic ambiguity in
integrated light studies from the uncertainty in determining the
fraction of light emitted by the various populations of stars, 
subsequent studies measured \ews\ in individual 
RGB stars (\markcite{dacosta89}Da Costa and Seitzer
1989, \markcite{olszewski91} Olszewski et al. 1991,
\markcite{suntzeff92} Suntzeff et al. 1992).  A method of ranking
globular clusters by \fe\, which is independent of both distance and
reddening was devised by \markcite{arm91}Armandroff and Da Costa (1991,
hereafter AD91), and adopted by almost all later practitioners
(\markcite{arm92}Armandroff et al. 1992 [ADZ92], \markcite{dacosta92}Da
Costa et al. 1992 [DAN92], \markcite{suntzeff93}Suntzeff et al. 1993
[S93], \markcite{geisler95}Geisler et al. 1995 [G95],
\markcite{dacosta95}Da Costa and Armandroff 1995 [DA95],
\markcite{suntzeff96}Suntzeff and Kraft 1996 [SK96],
\markcite{rutledge97}Paper I). 

AD91 fully describe the technique, but we summarize the salient
features.  The \ews\ is measured in probable member RGB stars as a
function of the V magnitude above the horizontal branch (HB).  Due to a
combination of both \logg\ and \teff\ decreasing as \vhbmv\ increases,
\ews\ increases going up the RGB with a slope of
$\sim0.64~\mAA$~mag$^{-1}$ (\markcite{rutledge97}Paper I).  It has been
shown empirically that this slope is independent of the cluster
metallicity (AD91, DA95, Paper I), and thus the \ews\ of a red giant star 
in any cluster can be adjusted to the level of the HB by
applying the correction $\ews_{HB} = \ews - 0.64 (\vhbmv)$.  The mean
value of $\ews_{HB}$ for all RGB stars measured in the cluster is then
denoted as the reduced equivalent width, \redew, which is taken to be a
calcium index for the entire cluster.  For the 52 globular
clusters we studied (Paper I), the \redew\ values ranged from 1.58~\AA\ for
NGC~4590 ($\fe_{ZW84} = -2.09$) to 5.41~\AA\ for NGC~6528 ($\fe_{ZW84}
= +0.12$).  With a typical $\sigma(\redew)$ of $\sim0.08~\mAA$, this
provides approximately 48 resolution elements  (each of $\sim0.05$ dex)
with which to rank the globular clusters in $\fe_{ZW84}$.

\section{Cluster Reduced \ca\ Equivalent Widths: \redew} \label{comp}

A large number of \ews\ measurements for globular cluster RGB stars are
now found in the literature. We have transformed these to our Paper I
system in two ways: the first involves transforming \redew\ values,
while the second involves transforming \ews\ values and then
calculating \redew\ from them.  As a result, 44 cluster \redew\ values
were transformed onto the Paper I system, 25 of which were in the Paper
I sample.  This increases the Paper I sample size from 52 to 71
clusters.  The transformed \redew\ values from the various authors,
along with our final adopted \redew\ value for each cluster, are
presented in Table~\ref{redew}, where the columns are, respectively, 1)
the consecutive cluster number; 2-4) the NGC, other cluster, and IAU
names; 5) the \redew\ values from Paper I; 6) the DA95 \redew\ values
transformed to the Paper I system, as described by the first method
below; 7) other \redew\ values transformed to the Paper I system, as
described by the second method below; 8) the final adopted \redew\ value
calculated as described below; and 9)  the reference from which the
\ews\ values were taken to calculate the \redew\ given in column 7.
The number following the $\pm$ sign in columns five through eight
represent our one sigma error estimates on these values.

DA95 transformed the \ews\ values of AD91, DAN92,
ADZ92, and S93 onto their system to calculate \redew\ values for
their calibrating globular clusters.  Using 13 globular 
clusters that we observed in common with DA95, we obtain the 
linear regression (see Figure~\ref{transform.grp}),
\begin{displaymath}
\redew_{Paper I} = 0.94 (\pm 0.02) \cdot \redew_{DA95} +0.06 (\pm 0.02) 
~~\me = 1.35,
\end{displaymath}
using an algorithm which allows for errors in both directions
(\markcite{stetson89ls}Stetson 1989).  As described in Paper I, the
\me\ value is a $\chi^2$ variable which measures the scatter about the
fit, where a value of one indicates that the scatter is consistent with
the observational errors, and larger values indicate that either the
errors have been underestimated or the relation is non-linear.   The
above regression was used to transform 24 \redew\ values from
DA95 onto our Paper I system (see Table~\ref{redew}), where the errors
tabulated include transformation errors.  We did not use NGC~5927 to
determine the regression due to the problems discussed by DA95\footnote
{Of the 14 NGC~5927 stars discussed by DA95, three were
contaminated by TiO, six had \ews\ values close to the 47~Tuc fiducial,
and five had \ews\ values $\sim 0.6~\mAA$ larger.  They used the five
larger values to define \redew\ for NGC~5927, and proposed that the
bifurcation is most likely a result of rather large errors
in the photometry they used (\markcite{menzies74}Menzies 1974) and/or
differential reddening across the cluster, as suggested by
\markcite{menzies74}Menzies and more recently by \markcite{cohen95}Cohen
and Sleeper (1995).  Unfortunately, the stars in our Paper I sample were
not observed in the new CCD photometry of NGC~5927
(\markcite{saranorris94}Sarajedini and Norris 1994), so we could not
test DA95's hypothesis.  Since the scatter of the 11 stars
about our fit to this cluster is consistent with our observational
errors, it seems that differential reddening among our stars was not
appreciable. (The stars from DA95 and Paper I were both selected to
avoid the region defined by \markcite{menzies74}Menzies (1974) to be
differentially reddened.) Our final \redew\ value for this cluster is
simply the weighted mean of the transformed DA95 value and the 
Paper I value.}.

For the remainder of the clusters in AD91, ADZ92, DAN92 and S93 that DA95 did
not use as calibrating clusters, and for the clusters observed by G95
and SK96, the following transformation technique was used.  The
individual \ews\ values were transformed onto the Paper I system using the
regressions given in Table~6 of Paper I, where the errors attached to each
value include the errors incurred from the transformation. The
\redew\ of each cluster was then calculated as the weighted mean of
$\ews\ - 0.64 (\vhbmv)$ for each star in the cluster, where the weights
were assigned as $1/\sigma^2$.  The errors for these \redew\ values,
listed in column 7 of Table~\ref{redew}, were calculated as 
$(\Sigma 1/(\sigma^2))^{-0.5}$ if the \me\ value was $\leq~1.0$, and as 
$\me \cdot (\Sigma 1/(\sigma^2))^{-0.5}$ if \me\ was $>~1.0$, where the
sum is taken over the number of stars observed in the cluster.  In
total, 21 cluster \redew\ values were transformed onto the Paper I
system using this technique.

The adopted \redew\ value for each cluster was calculated as a 
weighted mean of the individual \redew\ values listed in 
Table~\ref{redew}, where the weights were assigned as 
$1/\sigma^2$.  The errors were calculated as ($\Sigma
1/(\sigma^2))^{-0.5}$ if the \me\ value was $\leq~1.0$, and as $\me
\cdot (\Sigma 1/(\sigma^2))^{-0.5}$ if \me\ was $>~1.0$, where the sum
is taken over the available \redew\ values discussed above.  In order
to calculate the \redew\ index, it is necessary to measure the  V
magnitude of the HB, which introduces another source of error in
\redew. We have allowed for an uncertainty of $\pm 0.1$~mag in \vhb, by
adding in quadrature an additional error of $\pm 0.064~\mAA$ to the
error estimate described above.  This final error estimate is given in
column 8 of Table~\ref{redew}.

\placefigure{transform.grp}
\placetable{redew}

\section{The Zinn and West Relative Metallicity Scale } \label{relative}

The relative metallicities of globular clusters can be estimated by a
variety of techniques (\markcite{kraft79}Kraft 1979), most of which
measure either line blocking in the cluster integrated light, or some
metallicity sensitive parameter in the color-magnitude diagram (CMD),
while some of the more recent techniques employ equivalent width
measurements of strong spectral features.  The ZW84 \fe\ scale is a
relative ranking of 121 Galactic globular clusters placed on Cohen's
(see \markcite{frogel83}Frogel et al. 1983 for a compilation and
references) metallicity scale with a precision of $\sim$0.1 dex.  ZW84
combined a variety of relative ranking indices to increase their sample
size and average over various observational uncertainties.  Since the
ZW84 scale has been widely applied and contains the information present
in most relative ranking indices available at the time of its
compilation, we first focus our attention on it.

\subsection{\qzinn\ Ranking} \label{q39}

Seventy-nine globular clusters were ranked with the photometric \qzinn\ 
index (\markcite{zinn80b}Zinn 1980b), which is a filter system 
applied to integrated light to measure line blocking over $\sim$190
\AA\ centered on the \ca\ H and K lines.  Integrated light spectrograms
of 60 clusters were subsequently used to measure the pseudo-equivalent 
widths of the \ca\ K , the G band, and the \ion{Mg}{1} $b$ lines
(\markcite{zinn84}ZW84).  These rankings were transformed to
\qzinn\ values, which increased the number of clusters with
observed or inferred \qzinn\ indices to 74.  The weighted mean of \qzinn\ was taken 
by ZW84 to determine the spectroscopically augmented \qzinn\ values 
for clusters with more than one estimate.  
Except where specifically noted, we adopt the 
spectroscopically augmented \qzinn\ values of ZW84 for use throughout 
this paper; were the photometric \qzinn\ values used in the final analysis,
our results would not change significantly. 
Due to higher \teff\ values, blue HB stars produce great flux at
ultraviolet wavelengths.  Therefore, clusters with extremely blue HBs
are expected to have their \qzinn\ values diluted by this flux, and
thus their \fe\ values underestimated.  For the photometric \qzinn\ index, 
\markcite{manduca83}Manduca
(1983) provides theoretical support for this effect, while
\markcite{smith84} Smith (1984) and \markcite{frogel83} Frogel et al.
(1983) provide observational evidence.  However, ZW84 provide counter
arguments which suggest that their spectroscopically augmented \qzinn\ 
is only weakly dependent on HB morphology, if at all.

\placefigure{q39.grp}

In Figure~\ref{q39.grp}, we
plot each of these four \qzinn\ indices against \redew, where symbols represent
different ranges in \lee\ (\markcite{lee94}Lee et al. 1994).
The \qzinn\ values employed in the bottom right panel of Figure \ref{q39.grp}
are the purely photometrically determined \qzinn\ indices.
All four indices correlate well but non-linearly with \redew; indeed, all
suggest a break at \redew$\sim$3.5-4.0.
In the present sample, the second-parameter problem appears in its
full-blown form over a surprisingly narrow range of calcium-triplet
strength:  the calcium-weakest cluster with a predominantly red HB
[$\lee < 0$] is NGC~362, with $\redew = 3.72~\mAA$,
while the calcium-strongest cluster with a predominantly blue HB
[$\lee > 0$] is NGC~6266, with $\redew = 4.01~\mAA$.
This range of \redew\ corresponds approximately to --1.40 $\leq$
$\fe_{ZW84} \leq$ --1.25.
This is in contrast to some earlier
studies\footnote{
However, note that the present sample does
not include the northern clusters M3 = NGC~5272 and NGC~7006, which form a
famous second-parameter trend with M13 = NGC~6205 and M2 = NGC~7089 at a
metallicity near \fe = --1.6 on this scale.  Still, even M3 has a
``bluish'' HB in the sense of the present discussion, with
$\lee = 0.08$.
}
, e.g. \markcite{lee94}Lee, Demarque, and Zinn (1994), which found 
the second-parameter
effect to extend  over the range --2.0 $\ltsim$ \fe\ $\ltsim$ --1.1.
Therefore, the possibility exists that previous estimates of the {\it
extent\/} of the second-parameter problem may have been inflated slightly
by the inclusion of clusters with metallicities as uncertain as $\pm$0.2
or $\pm$0.3~dex.  Only the accumulation of a considerably larger sample of
highly precise metallicity indicators will delimit the true abundance
range of the second parameter effect.  However, within the restricted
range --1.40 $\ltsim$ $\fe_{ZW84} \ltsim$ --1.25 it is certainly true
that the second parameter problem is real, as clusters of all five
HB classes are found there:  our ``very blue'',
``blueish,'' ``reddish,'' and ``very red'' classes, as well as the
prototype of the ``bimodal HB'' class, NGC~2808.  Within the narrow range
3.72~\AA $\leq \redew \leq$ 4.01~\AA, it is evident from the bottom right panel
of Figure~\ref{q39.grp} that (within the small-number statistics of the present sample)
clusters with red HBs (closed symbols) tend to have larger values of the
photometric \qzinn\ index than those with blue HBs (open symbols) of
comparable \redew\ values.  Furthermore, if we extend the range under
consideration to 3.3~\AA $\ltsim \redew \ltsim$ 4.3~\AA, it appears that clusters
with blueish HBs (open diamonds) tend to lie above those with very blue
HBs (open triangles) and clusters with very red HBs (closed circles) tend
to lie above those with reddish HBs (closed squares).  Thus, the
photometric \qzinn\ index appears to increase systematically from clusters
with very blue, to blueish, to reddish, to very red HBs at $\fe_{ZW84}
\sim$ --1.3.  Given the present sample size, this result is hardly
definitive, but it does provide (weak) support to the contention of
\markcite{manduca83}Manduca (1983), 
\markcite{smith84}Smith (1984), and 
\markcite{frogel83}Frogel et al. (1983), that the
photometric \qzinn\ index is affected by differences in HB
morphology at fixed metal abundance.  However, little or no corresponding
trend of Ca II K-line strength, Mg I b strength, or G-band strength with
HB morphology is seen (the other three panels of Figure~\ref{q39.grp}).  
A more detailed discussion of the implications of these observations is 
given in the next section.

\subsection{Horizontal-Branch Morphology and $\fe_{\qzinn}$} \label{hb}

The relative ranking of the majority of the globular clusters in ZW84
has been determined either from 1) the \qzinn\ index, $\fe_{Q_{39}}$,
or, 2) the weighted mean of eight other independent ranking indices,
$\fe_{avg}$, or 3) the average of the two.  We have plotted the
\fe\ derived from each of the first two  methods, as well as the
difference between them (if both measures are available), as a function 
of \redew\ in Figure~\ref{fecomp.grp}.  Both \fe\ values show very 
good correlation with
\redew, but again, both are obviously non-linear with respect to
\redew.  The mean difference ($\fe_{Q_{39}} - \fe_{avg}$) for all
clusters in common is $-0.018 \pm 0.17 \mAA$.  This scatter is roughly
consistent with the expected observational errors, as noted by ZW84,
but there are a few features in the bottom panel that should be
emphasized.   

\placefigure{fecomp.grp}

At the metal-poor end, the \fe\ values are in good
agreement except for NGC~6397, for which $\fe_{Q_{39}}$ is $\sim 0.4$~dex 
lower than $\fe_{avg}$.  At the metal rich end, $\fe_{Q_{39}}$ is
slightly larger than $\fe_{avg}$ for every cluster except NGC~6352,
which only has a spectroscopic \qzinn\ value.  Excluding this cluster,
the mean value of $\Delta \fe$ for the five clusters with $\redew >
4.5~\mAA$, and $\lee \leq -0.5$ (\ie\ very red HBs) is 
$0.13 \pm 0.02~(s.d.m.)$~dex.  In
the range of \redew\ from 2.5 to 4.5 \AA, all of the clusters with
$\lee > 0.5$ (\ie\ very blue HBs) have $\Delta \fe$ values less than 
zero, except NGC~5286, and NGC~6218\footnote{
The clusters NGC 6266 and NGC 6626, which were used to analyze the 
dependence of the photometric \qzinn\ on HB type in \S \ref{q39}, are 
not present in this plot since they do not have $\fe_{avg}$ values. 
}.  Similarly, all clusters in this
range with $\lee \leq -0.5$ have $\Delta \fe$ values greater than zero,
except NGC~6712.  

From the studies of \S \ref{q39} and this section, we conclude that 
HB morphology does influence the \qzinn\ index, but, given the latter's 
significant observational uncertainties, HB
morphology likely has little influence on the overall $\fe_{ZW84}$ ranking. 
Accordingly, we have combined the ZW84 \fe\ ranking with our
\redew\ ranking to produce a relative metallicity scale of (presumably)
greater precision for the globular clusters in our sample.

\subsection{Most Probable cluster abundance on the Zinn and West Scale - $<\fe_{ZW84}>$} \label{zw84}

The ZW84 \fe\ values are plotted against \redew\ 
in Figure~\ref{zw84.grp}, where the symbols represent the different
techniques used to calculate $\fe_{ZW84}$.  In the upper panel, we plot 
those clusters which have $\fe_{ZW84}$ values which were calculated as
the weighted mean of $\fe_{Q_{39}}$ and $\fe_{avg}$.  This mean should
reduce the sensitivity of $\fe_{ZW84}$ to HB morphology,
as well as represent a well defined set of  \fe\ values.  We fit a
cubic polynomial allowing for errors in both directions,

\begin{displaymath}
\fe_{ZW84} = -3.005 + 0.941 \redew -0.312 \redew^2 + 0.0478 \redew^3
~~m.e.1 = 0.59,
\end{displaymath}
which allows $\fe_{ZW84}$ and \redew\ metallicity rankings to be merged
onto a common system.

NGC~1851
was not used in the regression due to its sensitive, and seemingly
anomolous position, near the upturn of the fit\footnote{
NGC~1851's anomolous position  was also noted by AD91 in their Figure 4.
}.
The exclusion of any other cluster from the regression (including
NGC~5927 and NGC~6218) does not significantly affect the fit.
Previous practitioners have generally calibrated their \redew\ index
to the ZW84 scale using two straight lines with
a break at $\redew \sim 3.75~\mAA$ (DA95, SK96).  We have transformed the
calibration lines of DA95 to the Paper I system and overplotted them
on our data in Figure~\ref{zw84.grp}.  For $\redew \lesssim 4.75~\mAA$, the
maximum deviation of our cubic fit from these calibration lines is
$\sim$ 0.05 dex. However, for \redew\ $>4.75~\mAA$, the deviations become
much larger, which no doubt reflects uncertainties in both quantities
for the metal-richer clusters.  The excellent agreement between our
cubic polynomial and DA95's two linear relations for 
clusters having \redew\ $<4.75~\mAA$ suggests both that we have adequately 
transformed DA95
\redew\ values onto our system, and that the final calibration with
$\fe_{ZW84}$ by either approach is indistinguishable within the
observational uncertainties. 

\placefigure{zw84.grp}
\placetable{feh}

In the bottom panel, all the clusters in our sample with $\fe_{ZW84}$
values are plotted.  It is apparent that much of the scatter in this
relation is caused by $\fe_{ZW84}$ values that were not derived from the weighted
mean of $\fe_{Q_{39}}$ and $\fe_{avg}$, which indicates that the predominant
source of the scatter is in $\fe_{ZW84}$.  In order to  
incorporate the information present in both scales, we have determined 
the most probable position of each cluster on the cubic polynomial.
This is done by calculating the point on the polynomial curve that 
penetrates most deeply into the nest of error ellipses centered on the 
observed position of each cluster (\markcite{stetson89ls}Stetson 1989).  
The results of this procedure are presented in Table~\ref{feh}, where the 
columns are respectively:
1,2) NGC and other names;
3) $\fe_{ZW84}$ as given by ZW84, with updates taken from 
\markcite{zinn85} Zinn (1985);
4,5) $<\redew>$,$<\fe_{ZW84}>$, which are the most probable values 
of these quantities calculated as described above;
6) $\fe_{CG97}$, which are the \fe\ values derived from transforming
our $<\redew>$ onto the high dispersion \fe\ scale of 
\markcite{carretta97} CG97 (see \S \ref{absolute});
and 7) notes for individual clusters.

The numbers after the $\pm$ sign in columns 4-6 represent our 
1 $\sigma$ error estimates on these quantities.  The error
estimates for $\fe_{CG97}$ are described in \S \ref{absolute},
while the error estimates for  $<\redew>$ and $<\fe_{ZW84}>$ were 
calculated as follows:

\begin{eqnarray}
\nonumber
\sigma_{<\fe_{ZW84}>}&=&K\cdot \left(\frac{1}{\sigma^{2}_{\fe_{ZW84}}} +
\frac{1}{m^{2\phm{-}}\cdot \sigma^{2}_{\redew}}\right)^{-0.5}, \\
\nonumber
\sigma_{<\redew>}&=&K\cdot \left(\frac{1}{m^{-2}\cdot 
\sigma^{2}_{\fe_{ZW84}}} +
\frac{1}{\sigma^{2}_{\redew}}\right)^{-0.5},
\end{eqnarray}
where $\sigma_{\fe_{ZW84}}$ is listed in column 3 of Table~\ref{feh},
$\sigma_{\redew}$ is listed in column 8 of Table~\ref{redew}, $m$ is 
the gradient of the cubic polynomial at $<\redew>$ in units of
dex\ts$\mAA^{-1}$, and $K$ is the ratio
of the deviation from the fit, 
$((\redew-<\redew>)^2 + (\fe_{ZW84}-<\fe_{ZW84}>)^2)^{0.5}$, to the
expected 1$\sigma$ deviation, 
$(\sigma^{2}_{\fe_{ZW84}}+ \sigma^{2}_{\redew})^{0.5}$.  $K$ is 
restricted to be $\geq 1$.  The typical precision with which these 
clusters can now be ranked on the ZW84 scale is $\sim \pm 0.05$ dex, which
is an improvement over the original ZW84 precision by a factor of $\sim$ 2.  
It should be emphasized that the 
$\sigma$ values attached to these quantities represent the precision with 
which the clusters can be ranked, and not the accuracies.  The absolute
metallicity scale is uncertain by at least 0.2 dex, as discussed 
in \S \ref{absolute}.

The metal-rich portion of the transformation is made uncertain by the lack of
high quality data.  The cubic 
polynomial that we have employed, and the linear fit employed by DA95
diverge significantly in this region, but the most probable positions
of the clusters do not change drastically depending on what 
calibration line is used.  For example, NGC~5927 has a most 
probable value on the cubic polynomial of $-0.32 \pm 0.08$ dex,
whereas on the DA95 line, it has a  most probable value $\sim -0.42$ dex.
The most probable value of NGC 6553 is $-0.18 \pm 0.12$ dex on the 
cubic fit, and $\sim -0.29$ dex on the DA95 line.  NGC 6528, which is not
listed in Table~\ref{feh} since it lies beyond the last calibrating
cluster, has a most 
probable value on the cubic polynomial of $+0.28 \pm 0.16$ dex,
whereas on the DA95 line, it has a most probable value $\sim 0.0$ dex.
Therefore, even though the calibration of the $\fe_{ZW84}$ values for 
the metal-rich clusters is uncertain, relative rankings appear to be 
robust.

\section{High Dispersion Metallicity Scales} \label{absolute}

Ultimately, the absolute \fe\ scale must 
be established through curve-of-growth analysis of 
high dispersion spectra (henceforth referred 
to as HDS) for many globular cluster stars.  However, due to the 
susceptibility of such analysis to systematic errors, which 
escalate rapidly as \fe\ increases, the absolute \fe\ scale is still the
subject of considerable controversy 
(\markcite{carretta97}CG97).  The debate over 
the metal-rich end of the metallicity scale in the early 1980's is 
well known (\markcite{freeman81}see, e.g., Freeman and Norris 1981), where  
photographic HDS with echelle spectrographs 
(\markcite{pila80}Pilachowski et al. 1980,
\markcite{cohen80}Cohen 1980) yielded significantly smaller
\fe\ values than previously suspected.  For this reason, the 
ZW84 scale is not based entirely on high-dispersion results, as 
discussed in \S\ref{linear}.

In an attempt to provide a modern high-dispersion calibration for
our \redew\ index,  we have culled from the literature all the recent
HDS \fe\ estimates from groups who have studied at least four of the
clusters in our sample.  These values are presented in 
Table~\ref{hds} where the 
columns are, respectively:
1,2) NGC and other names;
3) \fe\ values of Cohen used to calibrate the ZW84 scale;
and 4-8) the \fe\ values of five different HDS groups;
the references for each group are listed at the bottom of the
table.  For a given cluster, the range of \fe\ values from different
groups indicates that the absolute metallicity scale is still
uncertain by at least 0.2 dex. 

\placetable{hds}

We compare our calcium \redew\ index to the \fe\ values determined by
individual groups\footnote{ We assume that analyses done within a given
group make the same assumptions about \em gf \rm values, the absolute
solar Fe abundance, \teff\ scale, etc.} in Figure~\ref{hds.grp}, where
the solid line overplotted on the data is the cubic polynomial fit to
the $\fe_{ZW84}$ data.  The \fe\ values of Cohen show the same
non-linear relationship to \redew\ as $\fe_{ZW84}$, since the ZW84
scale was calibrated with her data.  However, none of the other HDS
\fe\ results show any strong evidence for a non-linear relationship
with \redew.  This corroborates the suggestion of
\markcite{carretta97}CG97 that the ZW84 \fe\ scale may be non-linear
with respect to the true \fe\ scale.

\placefigure{hds.grp}

The \fe\ scale defined by CG97 is based upon the largest, most self-consistent
HDS analysis that presently exists for the globular clusters. 
They have reanalyzed high quality equivalent widths from several different
sources, including GO89, M93, and SKPL, using a
homogeneous compilation of stellar atmosphere parameters, \em gf \rm 
values, and the Kurucz (1992) stellar atmospheres.  
A linear fit which allows for errors in both directions yields,
\begin{displaymath}
\fe_{CG97} = -2.66 (\pm 0.08) + 0.42 (\pm 0.02) \cdot  <\redew> ~~m.e.1 = 1.51. 
\end{displaymath}
This relation is plotted as a dotted line in Figure~\ref{hds.grp}, and
is used to transform the $<\redew>$ values into $\fe_{CG97}$ in 
Table~\ref{feh}. The errors have simply been calculated
as,
\begin{displaymath}
\sigma_{\fe_{CG97}} = m \cdot \sigma_{<\redew>},
\end{displaymath}
where m is the slope of the above linear relation in units of 
dex\ts$\mAA^{-1}$, and $\sigma_{<\redew>}$ is listed in 
Table~\ref{feh}.  We have not included the transformation
errors since we are only interested in how the clusters rank on this
scale.  As stated in \S \ref{zw84}, $\sigma_{\fe_{CG97}}$
represents the precision with which the cluster can be ranked, and
not the accuracy.  For the calcium index, we chose to use the most 
probable value, $<\redew>$, rather than the measured value, \redew,
since the technique used in \S\ref{zw84} to calculate
 $<\redew>$ adds information from the various ranking systems upon
which the ZW84 scale is based and, thus, is likely to improve the 
precision of relative ranking.  However, for any cluster whose
$\fe_{CG97}$ value changes by more than 0.05 dex when \redew\ is 
used in the calibration instead of $<\redew>$, we list what 
$\fe_{CG97}$ would be in a footnote to Table~2.

\section{Discussion} \label{discuss}

The foregoing determination of globular cluster \redew\ indices appears
to have increased by a factor of $\sim 2$ the precision with which we
can rank globular clusters by metallicity on a scale based upon
$\fe_{ZW84}$. Such an improvement would appear to support the
effectiveness of the \ca\ triplet as a metallicity indicator,
especially in the context of AD91's \redew\ index for old, metal-poor
stellar systems. In spite of our apparent success, we feel rather
sobered by issues surrounding the metallicity scale and the widespread
use of the \ca\ triplet as a surrogate for overall metal abundance,
as we briefly discuss here.

\subsection{\redew\ and Other Scales} \label{linear}

As shown above, \redew\ is tightly correlated with \fe\ for the
globular clusters we have studied.  Disturbingly, however, a
significantly non-linear correlation with $\fe_{ZW84}$ contrasts
sharply with the linear correlation we find with $\fe_{CG97}$. We have 
no particular \em a priori \rm reason to expect how \redew, an
empirical line-strength measure, should be related to \fe, and it
is unclear which, if either, \fe\ scale to trust.

The sense of the non-linear relation between \redew\ and $\fe_{ZW84}$
is that the abundance sensitivity of \redew\ decreases as the abundance
on the $\fe_{ZW84}$ scale increases (AD91, DAN92, S93, DA95,
SK96).  Most of the preceding  authors chose to transform their
\redew\ values into $\fe_{ZW84}$ via two linear relations with a break
at $\fe_{ZW84} \sim -1.5$.  Three potential reasons for the decrease in
abundance sensitivity of \redew\ with increasing abundance were
outlined by AD91, but to our knowledge this issue has not been
significantly explored since.  AD91 proposed that: i) weak metal lines
and molecular bands of TiO may be depressing the pseudo-continuum in
the more metal-rich clusters, thereby resulting in smaller measured
\ews\ values;  ii) at higher abundances, the increase in the strength
of the \ca\ lines occurs mostly in the wings, where it is harder to
measure than in the line cores\footnote{Note, however, that it is likely
that all of our \ca\ triplet measures are on the square-root portion of the
curve of growth (A. Irwin, private communication), hence AD91's suggestion
may be incorrect.}; and finally, iii) \cafe\ actually does
decrease with increasing \fe.

There are at least two other effects that should be considered when
trying to understand this nonlinearity.  Firstly, as noted by AZ88, \redew\
increases with \fe\ both because the abundance of Ca in the atmosphere 
is larger, and because \logg\ at the level of the HB is smaller.
Furthermore, an increase in cluster \fe\ causes \teff\ at the level of
the HB to be smaller. Despite the earlier studies, \markcite{jorg92}
Jorgensen et al. (1992) have shown through synthetic spectra
calculations that \ews\ has a relatively large and complex dependence
on \teff\ in the globular cluster metallicity range. It is clear that
\redew\ is sensitive to more than just a simple abundance increase in
globular clusters as their RGBs shift in the \logg, \teff\ 
plane.

Secondly, \redew\ may appear to lose sensitivity to \fe\ for the higher
metallicity clusters because all comparisons, so far, have been made
between \redew\ and $\fe_{ZW84}$, which may itself be non-linear with
respect to the true \fe\ (\markcite{carretta97}CG97 and below).  
The relative
indices used by ZW84 were placed on the \fe\ scale compiled by
\markcite{frogel83}Frogel et al. (1983).  For $\fe \leq -1.5$, the
scale is based on the early high-dispersion photographic spectroscopy
of individual giant stars by Cohen (\markcite{cohen78b}1978b,
\markcite{cohen79b}1979b, \markcite{cohen81}1981) and
\markcite{cohen82}Cohen and Frogel (1982), while for the more
metal-rich clusters, it is based on models of line blocking in
low-dispersion scans of individual stars (Cohen \markcite{cohen82}1982,
\markcite{cohen83}1983).  It is interesting that the ``break'' in
linearity between \redew\ and $\fe_{ZW84}$ also occurs at $\fe_{ZW84}
\sim -1.5$.  In addition, if only clusters with $\fe_{ZW84} \lesssim
-1.5$ are used to compare $\fe_{ZW84}$ and $\fe_{CG97}$, there is
very little evidence for a non-linear relationship between them. 

In summary, possible reasons for the non-linear relationship between
\redew\ and $\fe_{ZW84}$ can be split into three classes:  i) it
is a result of how \logg\ and \teff\ (and possibly turbulent velocity)
of the RGB stars vary for clusters
of different \fe\, and, thus, how the line formation physics behaves
for those variables;  ii) it is a result of Galactic formation
physics, where the more metal-poor clusters were enriched by different
mechanisms than were the metal-rich clusters, thereby producing
different \cafe\ ratios as a function of \fe\ (see below); or, iii)
$\fe_{ZW84}$ is non-linear with respect to the true \fe.  The
non-linear relationship may well result from  some combination of the
preceding factors.

Our referee suggested that further insight might result from plotting
\cah\ vs \redew. To investigate this suggestion, we employed the abundance
data in \markcite{carney96}Carney's (1996) critical compilation, 
where \cafe\ as well as $\fe_{zinn}$ and $\fe_{spec}$ are given. 
The former \fe\ from \markcite{zinn85}Zinn (1985) is essentially 
the $\fe_{ZW84}$ employed throughout this paper, while $\fe_{spec}$ 
was computed by Carney from spectroscopic abundance determinations in 
the literature, which he put onto a common scale of solar iron abundance.  
For the 18 clusters in common, we have computed 
$\cah_{zinn} = \fe_{zinn} + \cafe_{carney}$ and 
$\cah_{spec} = \fe_{spec} + \cafe_{carney}$, 
where quantities on the rights hand side are all from
Carney's review. These are shown in Figure \ref{cah.grp}, top and bottom,
respectively.  The linear regressions 
($\cah_{zinn}=0.35\redew - 2.36$,
$\cah_{spec}=0.32\redew - 2.23$), as well as the cubic relation found in
\S\ref{zw84} between \redew\ and $\fe_{ZW84}$, are plotted in each panel. 
The dispersion about the linear regression of \redew\ on $\cah_{spec}$ 
is about half that of $\cah_{zinn}$, whose data suggest a quadratic or 
higher fit would be more appropriate.  Moreover, on the \redew\ interval 
in common with \markcite{norris96}Norris, et al.'s (1996) Figure 7 for 
stars in $\omega$~Cen, the agreement of their slope and intercept is 
within the uncertainties of our Figure \ref{cah.grp} linear regressions.

From Figure \ref{cah.grp}, $\cah_{spec}$ would seem to be linearly correlated 
with \redew\ for these clusters. In turn, this would seem to imply that 
the non-linear relationship between \redew\ and $\fe_{ZW84}$ is not due
predominantly to line formation physics. These initial impressions
might be taken with some caution, however, because of (a) the caveats
raised in Carney's analysis, (b) the relatively small number of clusters
in common, and (c) concerns about the unknown role of systematic effects 
in high dispersion analyses as a function of \fe\ (and from one
investigator to another).

\subsection{Galactic Formation and the $\alpha$ Elements}\label{form}

The foregoing raises important questions of what \redew\ is actually
measuring over the parameter range of the globular cluster RGB stars,
and what it can tell us about the formation of the Milky Way.  We
briefly comment upon the second query here, and upon the first in the
next section.

\markcite{carney96}Carney (1996) has carefully reviewed spectroscopic
determinations of [$\alpha$/Fe] in globular clusters. He concluded
that the mean [$\alpha$/Fe] values determined from silicon and titanium do not
appear to vary over the range of [Fe/H]$=-$2.24 to $-$0.58, nor does
[O/Fe] for those RGB stars which are presumed to be unmixed, and they
are all strongly enhanced relative to solar values.  That the
$\alpha$ elements appear to share a common, uniform enhancement among
the globular clusters contrasts with the halo field star situation, in
which [O/Fe] and [$\alpha$/Fe] are enhanced by $+$0.3 to 0.5 dex for
\fe\ $ \lesssim -1.4\pm 0.3$ and then drop to solar values (e.g.,
Greenstein \markcite{green70}1970, Wheeler, et al. \markcite{wheeler89}
1989), in ways reminiscent of the relation between \redew\ and
$\fe_{ZW84}$ in Figure~4.  Interestingly, among the $\alpha$ elements
in globular cluster stars, [Ca/Fe] may behave more like [$\alpha$/Fe]
in the field halo stars; however,
Carney suggests that the observed decline of [Ca/Fe] as [Fe/H]
increases may be an artifact arising from the use of neutral calcium
lines to determine [Ca/Fe] in the available spectroscopic studies.

If, as Carney (1996) discusses, his interpretation of the fairly
extensive HDS data available for globular clusters is correct, it
suggests that, contrary to evidence from field halo star
studies, contributions from supernovae of Type Ia are absent from
the chemical evolution of the globular clusters or operated on a very
different time scale than commonly thought.  This conclusion doubtless
will be controversial, in part because of the contrasting field halo
star patterns and in part because arguments exist for a smaller age
range among the globular clusters than he advocates (see, e.g., 
\markcite{hesser95}Hesser 1995, 
\markcite{richer96}Richer, et al. 1996,
\markcite{stetson96}Stetson, VandenBerg \& Bolte 1996, 
\markcite{hesser96b}Hesser, et al. 1996b).  Nonetheless, there is no doubt that the
distribution of the $\alpha$ elements provides a clue, perhaps critical,
regarding how the Galactic halo formed and, thus, it is crucial to
understand the limitations of the \ca\ triplet as a metallicity
indicator.

\subsection{The Calcium Triplet as a Metallicity
Surrogate}\label{surrogate}

As reviewed in \S\ref{indicator}, the \ca\  triplet has not always
enjoyed its current popularity as a relatively easily determined
surrogate for metallicity.  For the Galactic globular clusters, others
have demonstrated, and we have confirmed and extended, that the triplet
appears to be a sensitive metallicity indicator.  However, as noted in
the preceding section, Carney (1996) raises concerns whether calcium
participates in the uniform enhancement patterns exhibited by O, Si and Ti. 
If the Galactic halo formed by mergers of smaller galaxies that had undergone
independent chemical evolution, element ratios could differ from one
set of clusters to another according to the history of supernovae and
the mixing of their ejecta in the galaxy of cluster origin.  

For our own efforts to determine ages of the far outer halo clusters
from Hubble Space Telescope photometry (cf.  Hesser, et al.  1996a
\markcite{hesser96a}, Stetson, et al.  1996\markcite{stetaas}), detailed
spectroscopic abundances will be quite challenging to obtain from
high-dispersion spectra; consequently, high reliance is currently
placed on the \ca\ triplet as a metallicity surrogate. Should the outer
halo clusters by and large represent a different family, it is
possible that calcium (and the $\alpha$ elements in general) may
have followed a different enrichment path than clusters in the inner
halo. Similar concerns might apply for other possible families of clusters
throughout the Milky Way. Although little evidence for deviations from the 
Figure \ref{zw84.grp} mean relation is evident for some of the suggested 
subsystems of Galactic globular clusters, \markcite{brown97}Brown
et al. (1997) find from spectroscopic analysis of Rup 106 and Pal 12 
that the $\alpha$-elements are \em not \rm enhanced over the solar ratio.
At least when interpreting \ca\ triplet data
for an individual globular cluster, we have strong evidence from
color-magnitude diagrams that an internal age range does not factor into
its abundance sensitivity.  The same cannot be said when interpreting
the results of \ca\ triplet measurements in dwarf spheroidal galaxies
(or other resolvable  stellar populations where a range of ages and
abundances are likely), nor, especially, when interpreting triplet
strengths in composite light spectra.  In such cases, additional
observational constraints on the parameters that affect the formation
of the triplet lines are required for reliable metallicities to be
derived from them.

It seems particularly unfortunate that, in spite of the heavy reliance
placed upon the calcium triplet in modern studies that strive to constrain
the formation of the Galaxy through metallicity and age determinations
of objects throughout the halo, no thorough modelling of the triplet
lines has been carried out in a way that would help resolve some of the
issues raised herein or by previous workers.  To our
knowledge, the most comprehensive theoretical work on the \ca\ triplet
has been  done by \markcite{jorg92}Jorgensen at al. (1992), but they
only include calculations for $\fe \geq -1$, and they do not include
the effects of line blanketing, which could play a significant role in
depressing the pseudo-continuum for cool RGB stars in the more metal
rich clusters.  To help rectify this situation, we have begun some
exploratory theoretical studies of the line formation physics for the
\ca\ triplet in collaboration with Ana Larson and Alan Irwin at the
University of Victoria.  We are theoretically simulating the AD91
\redew\ technique.  Among our goals is to determine if a non-linear
relationship between \redew\ and \fe\ is expected from the line
formation physics alone.  The oxygen enhanced isochrones of
\markcite{berg92}Bergbusch and Vandenberg (1992) are used to define the
\logg\ - \teff\ loci of RGB stars in globular clusters with
\fe\ ranging from 0 to $-2$.  Full synthetic spectra calculations are
then done using the Ssynth code (\markcite{larson96}Larson and Irwin,
1996), and a reduction technique identical to that described in Paper I
is applied to obtain \redew\ values. 

We strongly encourage other groups with the appropriate theoretical
tools to examine independently this empirical diagnostic, whose
relevance to many areas of Galactic and extragalactic astrophysics is
growing steadily.

\subsection{Final Remarks}\label{final}

The globular cluster \fe\ scale is fundamental to our understanding of
the age and chemical evolution of our Galaxy.  For more than a decade,
researchers have relied on the \fe\ scale of ZW84, which seems generally
to provide adequate relative rankings (see \S \ref{relative}), but
depends on the non-homogeneous \fe\ scale of \markcite{frogel83}Frogel
et al. (1983) for an absolute calibration (see \S \ref{indicator}).
Although the scale of \markcite{frogel83}Frogel et al. (1983) was state
of the art at the time, some of their data and analysis techniques are
almost twenty years old, and deserve revisiting using subsequent
improvements.  The homogeneous analysis of good quality,
high-dispersion spectra by CG97 appears to be an important first step
in defining a modern high dispersion \fe\ scale.  However, the CG97
\fe\ scale creates a paradox, since it is unclear at the moment whether
$\fe_{ZW84}$ or $\fe_{CG97}$ approximates the true \fe\ scale more
closely.  Since this issue is not likely to be resolved in the
near future, we have transformed our \redew\ values into \fe\ on both
scales (see Table~\ref{feh}) so that  researchers can perform their
analysis using either.

\pagebreak[0]
\acknowledgments

G.A.R. acknowledges the valuable computer assistance offered by Daniel
Durand, Gerry Justice, and Wes Fisher over the course of this project,
as well as the National Research Council of Canada for an opportunity to
work at DAO.  We thank the anonymous referee for a helpful report.

\clearpage
\begin{table}
\dummytable \label{redew}
\end{table}

\begin{table}
\dummytable \label{feh}
\end{table}

\begin{table}
\dummytable \label{hds}
\end{table}



\clearpage
\figcaption[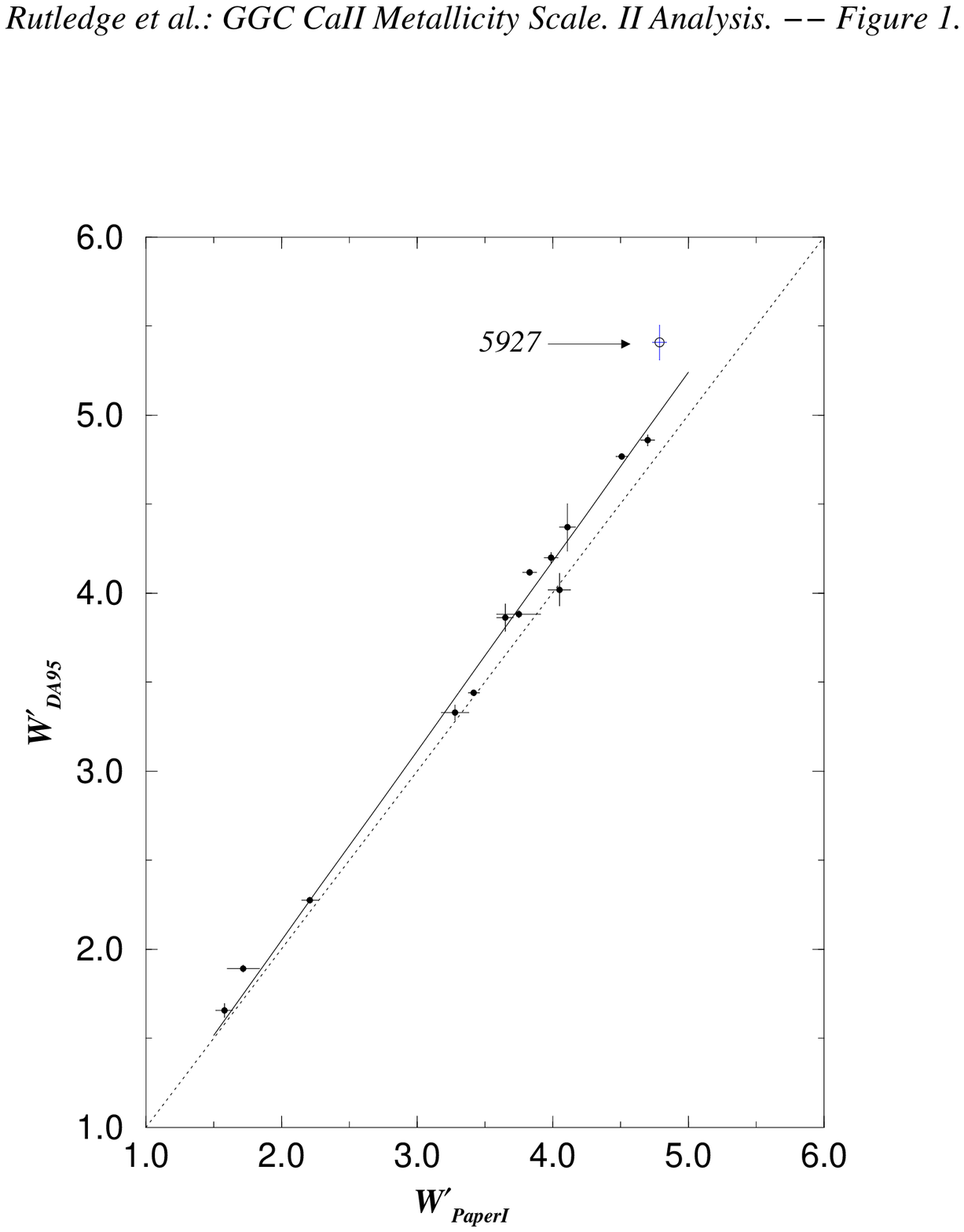]{The linear regression used to convert
the \redew\ values of DA95 into \redew\ values on the Paper I system.  The
cluster NGC 5927, plotted as an open circle, was not used to determine
the regression.  \label{transform.grp}}

\figcaption[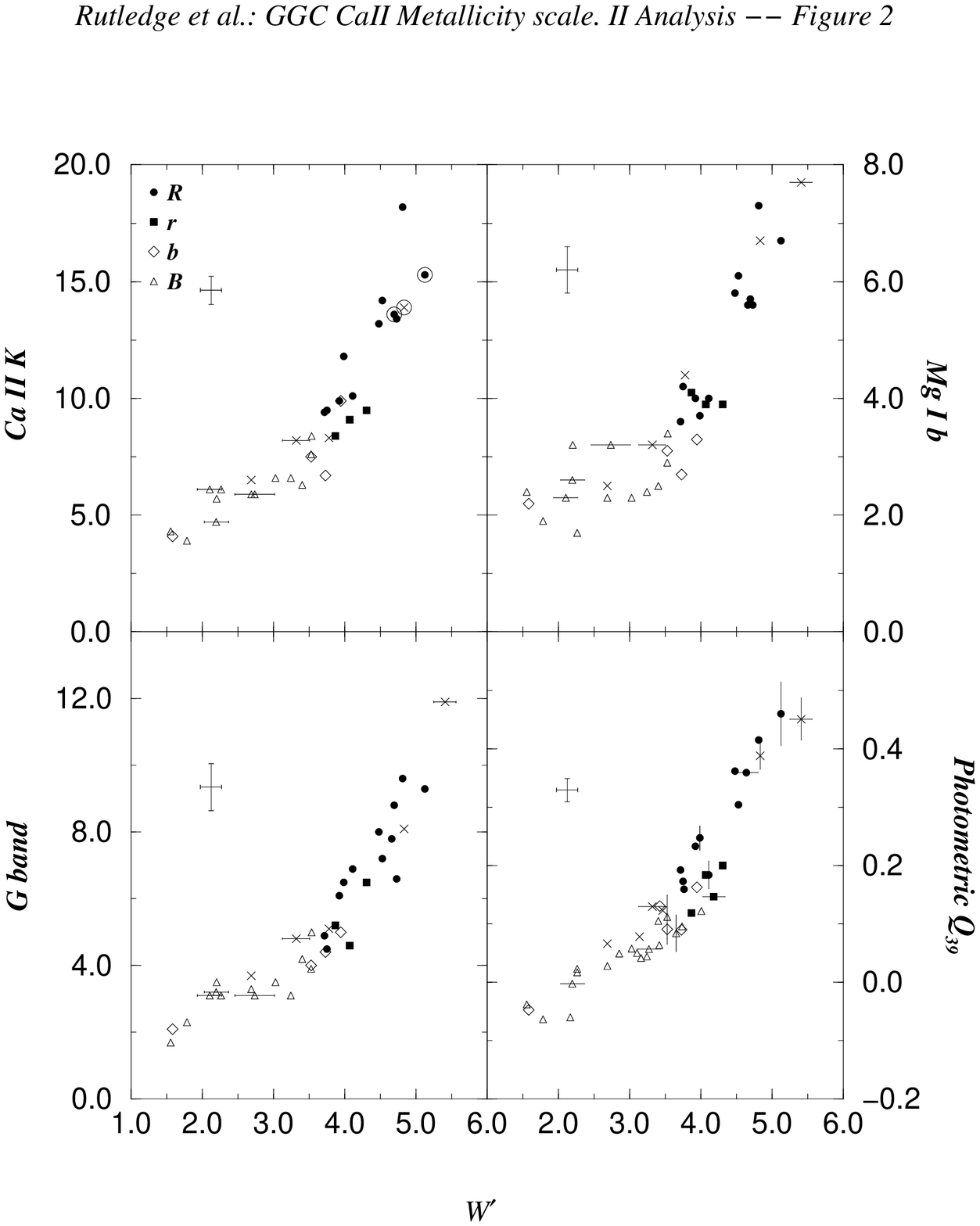]{A comparison of the four indices used by
ZW84 to construct their spectroscopically augmented $Q_{39}$ index.  Note that 
the bottom right panel is the original photometric \qzinn\ of Zinn (1980b).
Typical error bars are given in the upper
left corner of each plot.  An error bar is attached to an individual 
point only if its uncertainity  exceeds the typical value.  The different symbols
denote Lee et al. (1994) HB types as follows:  
\em solid circles \rm (R): $\lee \leq -0.5 $;
\em solid squares \rm (r): $-0.5 < \lee \leq 0.0$;
\em open diamonds \rm (b): $0.0 < \lee \leq 0.5$;
\em open triangles \rm (B): $0.5 < \lee $;
$\times$: \rm \lee\ is undefined.  
The letters R, r, b, and B are used in the figure legend as a reminder of 
the HB type.
The \em open circles \rm 
in the upper left plot are overlaid on clusters for which the  \ca\ K
equivalent widths have less than average weight (as defined by ZW84).
\label{q39.grp}}

\figcaption[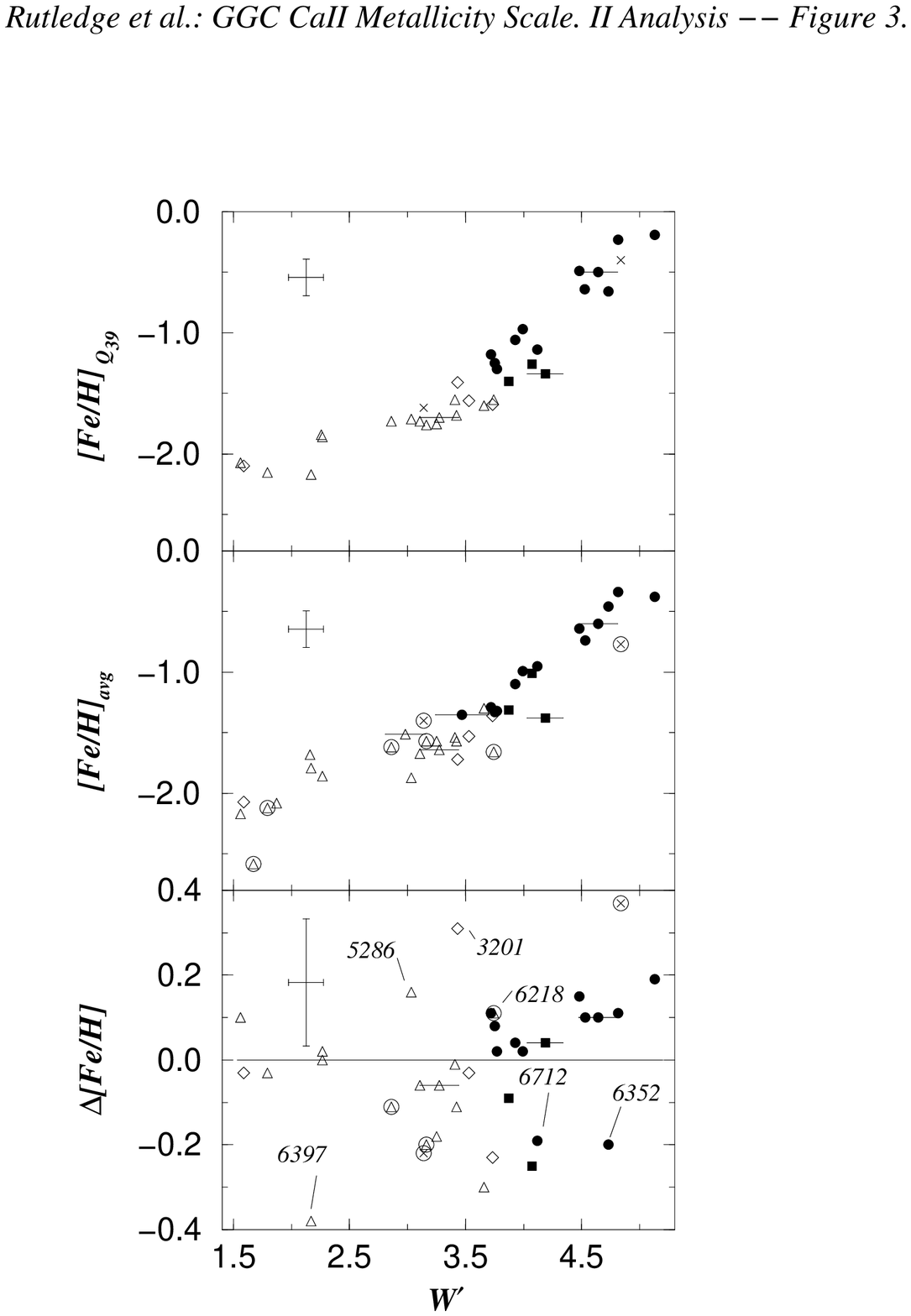]{A comparison of the spectroscopically augmented
$\fe_{Q_{39}}$ defined
by ZW84, which is a combination of photometric $Q_{39}$ and
transformed, low-dispersion spectral indices (upper panel), and
$\fe_{avg}$ defined using the average of an assortment of indices
(middle panel).  The bottom panel shows the difference, $\fe_{Q_{39}} -
\fe_{avg}$.  All \fe\ values have been taken from Table~5 of ZW84.  The
symbols are defined as in Figure~\ref{q39.grp}, except for the \em open
circles \rm which, in this plot, are overlaid on  clusters for which
$\fe_{avg}$ was obtained using only one index.  \label{fecomp.grp}}

\figcaption[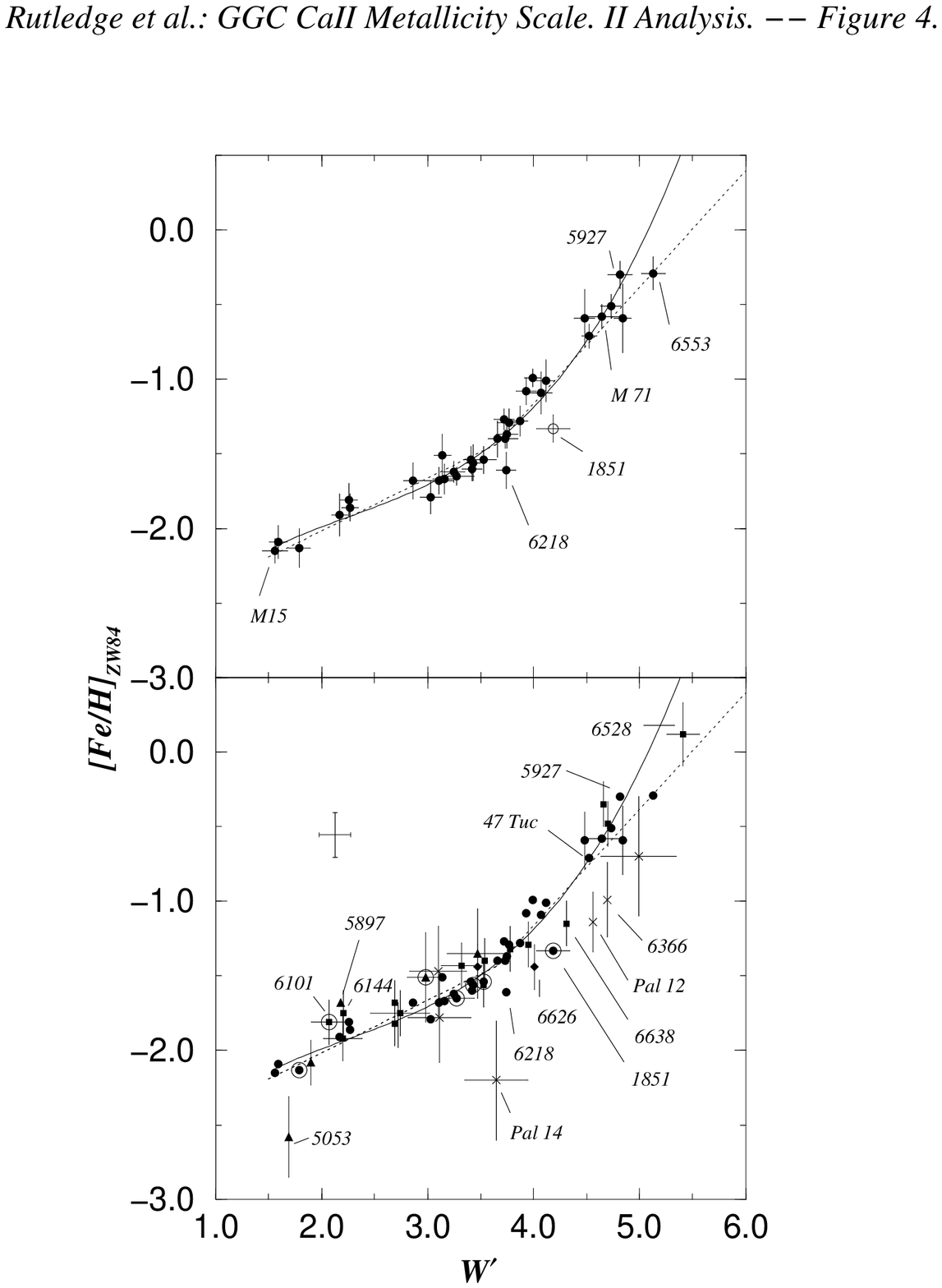]{ The \fe\ values from Table~6 of ZW84
have been plotted against our calcium index, \redew.  The upper panel
contains only clusters for which $\fe_{ZW84}$ was derived from a combination
of $\fe_{Q_{39}}$ and $\fe_{avg}$ (see Figure~\ref{fecomp.grp}).  These clusters
were used to fit a cubic polynomial, plotted as a solid line, which is then used
to define the most probable value of \fe\ and \redew\ for each cluster 
on the ZW84 scale. NGC 1851, plotted as an open circle, was not used in the
regression.  The dotted line, which abruptly changes slope at \redew\ = 3.64 \AA,
is the calibrating line of DA95 transformed to our Paper I system.
The lower panel shows all clusters that are in common to the two studies, where the 
symbols are defined as follows:
\em solid circles \rm when  $\fe_{ZW84}$ was derived from a combination
of $\fe_{Q_{39}}$ and $\fe_{avg}$;
\em squares \rm when  $\fe_{ZW84}$ was derived from $\fe_{Q_{39}}$
only;
\em diamonds \rm when $\fe_{ZW84}$ was derived from $\fe_{Q_{39}}$
only, where $Q_{39}$ was calculated using only the photometric $Q_{39}$ index;
\em triangles \rm when  $\fe_{ZW84}$ was derived from $\fe_{avg}$
only;
$\times$ \rm when  $\fe_{ZW84}$ was derived from some other, 
presumably less reliable, index.  The \em open circles \rm are overlaid on
clusters which van den Bergh (1993) defines as having retrograde
motions about the Galaxy.  The error bars in the upper left corner
represent errors of $\pm 0.15$~dex in $\fe_{ZW84}$, and $\pm 0.15~\mAA$
in \redew.  Error bars are attached to the symbols only if they exceed
these values.  The solid and dotted lines are reproduced from the top
panel.  \label{zw84.grp}}

\figcaption[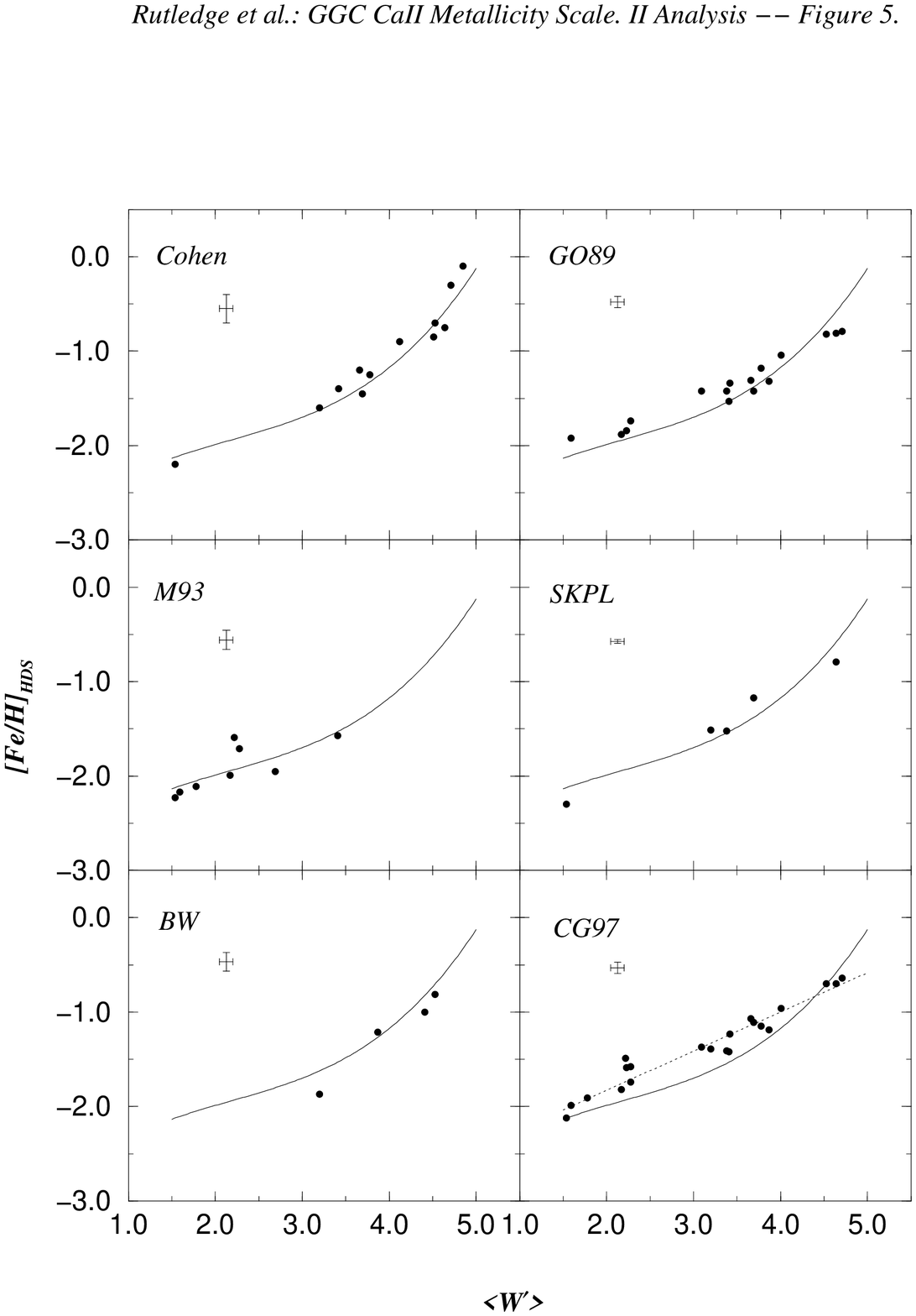]{
The \fe\ values from Cohen, and from five more recent high-dispersion spectroscopy 
studies, are compared to the most probable $<\redew>$ index listed in Table~\ref{feh}.  
The error
bars in the upper left corner give the typical errors for 
each of the quantities. The solid
line in each panel is the cubic fit to $\fe_{ZW84}$ from 
Figure \ref{zw84.grp}.  The dotted line in the CG97 panel 
shows the linear regression used to transform the $<\redew>$ values
to $\fe_{CG97}$.  \label{hds.grp}}

\figcaption[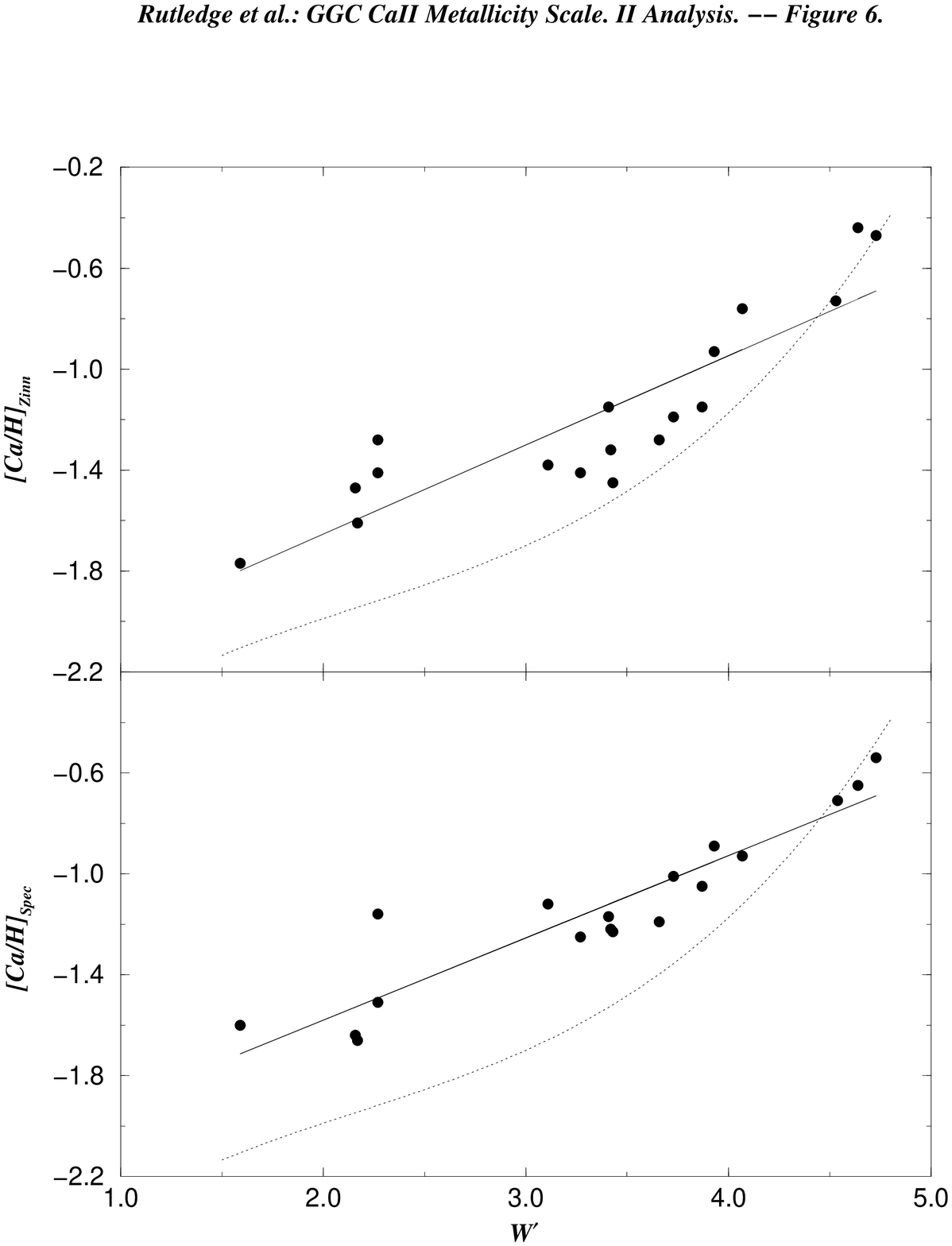]{
A comparison of \cah\ with our \redew\ measurements; see text for details.
\label{cah.grp}}

\begin{references}
\reference{alloin89} Alloin, D., and Bica, E. 1989, \aap, 217, 57
%
\reference{anderson74} Anderson, C.M. 1974, \apj, 190, 585
%
\reference{arm89} Armadroff, T.E. 1989, \aj, 97,375
%
\reference{armzinn88} Armandroff, T.E., and Zinn, R. 1988, \aj, 96, 92 (AZ88)
%
\reference{arm91} Armandroff, T.E., and Da Costa, G.S. 1991, \aj, 101, 
1329 (AD91)
%
\reference{arm92} Armandroff, T.E., Da Costa, G.S., and Zinn, R.  1992, 
\aj, 104, 164 (ADZ92)
%
\reference{brown91} Brown, J. A., Wallerstein, G., and Oke, J.B. 1991, 
\aj, 101, 1693
%
\reference{brown92} Brown, J. A., and Wallerstein, G. 1992, \aj, 104, 1818
%
\reference{brown96} Brown, J. A., Wallerstein, G., and Zucker, D. 1996,
Formation of the Galactic Halo \ldots Inside and Out, ASP Conference
Series, Vol. 92, eds. Heather Morrison and Ata Sarajedini (San Francisco,
ASP), p. 355
%
\reference{brown97} Brown, J. A., Wallerstein, G., and Zucker, D. 1997,
\aj, in press
%
\reference{berg92} Bergbusch, P.A., and Vandenberg, D.A. 1992, \apjs, 81, 163
%
\reference{carney96} Carney, B.W. 1996, \pasp, 108, 900
%
\reference{carretta97} Carretta, E., and Gratton, R.G. 1997,
\aaps, 121, 95 (CG97)
%
\reference{cohen78} Cohen, J.G. 1978, \apj, 221, 788
%
\reference{cohen78b} Cohen, J.G. 1978b, \apj, 223, 487
%
\reference{cohen79} Cohen, J.G. 1979, \apj, 228, 405
%
\reference{cohen79b} Cohen, J.G. 1979b, \apj, 231, 751
%
\reference{cohen80} Cohen, J.G. 1980, \apj, 241, 981
%
\reference{cohen81} Cohen, J.G. 1981, \apj, 247, 869
%
\reference{cohen82} Cohen, J.G. 1982, \apj, 258, 143
%
\reference{cohen83} Cohen, J.G. 1983, \apj, 270, 654
%
\reference{cohen95} Cohen, J.G., and Sleeper, C. 1995, \aj, 109, 242
%
\reference{cohenfrogel82} Cohen, J.G., and Frogel, J.A. 1982, \apjl, 255 L39
%
\reference{dacosta95} Da Costa, G.S., and Armandroff, T.E. 1995, \aj,
109, 2533 (DA95)
%
\reference{dacosta92} Da Costa, G.S., Armandroff, T.E. , and
Norris, J.E. 1992, \aj, 104, 154 (DAN92)
%
\reference{dacosta89} Da Costa, G.S., and Seitzer, P. 1989, \aj, 97, 405
%
\reference{diaz89} Diaz, A.I., Terlevich, E., and Terlevich, R.
1989, \mnras, 239, 325
%
\reference{freeman81} Freeman, K.C., and Norris, J. 1981, \araa, 19, 319
%
\reference{frogel83} Frogel, J.A., Cohen, J.G., Persson, S.E. 1983,
\apj, 275, 773
%
\reference{geisler95} Geisler, D., Piatti, A.E., Clari\'a, J.,
and Minniti, D. 1995, \aj, 109, 605 (G95)
%
\reference{gratton89} Gratton, R.G., and Ortolani, S. 1989, \aap,
211, 41 (GO89)
%
\reference{green70} Greenstein, J.L. 1970, Comm. Astrophys. Space Sci., 2, 85
%
\reference{harris94} Harris, W.E. 1994, electronically published catalog,
McMaster University
%
\reference{hesser95} Hesser, J.E. 1995, in Stellar Populations, eds.
G. Gilmore, P. Van der Kruit (Dordrecht, Kluwer), p. 51
%
\reference{hesser96a} Hesser, J.E., Stetson, P.B., McClure, R.D., van den
Bergh, S., Harris, W.E., Bolte, M., VandenBerg, D.A., Bond, H.E., Fahlman,
G.G., Richer, H.B., Bell, R.A. 1996a, \baas, 28, 1363.
%
\reference{hesser96b} Hesser, J.E., Stetson, P.B., Harris, W.E., Bolte,
M., Smecker-Hane, T.A., VandenBerg, D.A., Bell, R.A., Bond, H.E., van
den Bergh, S., McClure, R.D., Fahlman, G.G., and Richer, H.B. 1996b, 
J. Korean Astron. Soc., 29, S111.
%
\reference{jones84} Jones, J.E., Alloin, D.M., and Jones, B.J.T. 
1984, \apj, 184, 787
%
\reference{jorg92} Jorgensen, U.G., Carlsson, M, and Johnson, H.R.
1992, \aap, 254, 258
%
\reference{kraft79} Kraft, R.P. 1979, \araa, 17, 309
%
\reference{kraft93} Kraft, R.P., Sneden, C., Langer, G.E., Shetrone, M.D.
1993, \aj, 106, 1490
%
\reference{kraft95} Kraft, R.P., Sneden, C., Langer, G.E., Shetrone, M.D.,
Bolte, M. 1995, \aj, 109, 2586
%
\reference{kurucz92} Kurucz, R.L. 1992, private communication
%
\reference{larson96} Larson, A.M. and Irwin, A.W. 1996, \aaps, 117, 189L
%
\reference{lee94} Lee, Y-W, Demarque, P., and Zinn, R. 1994, \apj, 423, 248
%
\reference{manduca83} Manduca, A. 1983, \baas, 15, 647.
%
\reference{menzies74} Menzies, J. 1974, \mnras, 169, 79
%
\reference{minniti93} Minniti, D., Geisler, D., Peterson, R.C., Claria, J.J.
1993, \apj, 413, 548 (M93)
%
\reference{norris96} Norris, J.E., Freeman, K.C. and Mighell, K.J. 1996,
\apj, 462, 241
%
\reference{olszewski91} Olszewski, E.W., Schommer, R.A., Suntzeff, N.B.,
and Harris, H. 1991, \aj, 101, 515
%
\reference{pila80} Pilachowski, C.A., Canterna, R., Wallerstein, G. 1980,
\apjl, 235, L21
%
\reference{richer96} Richer, H.B. et al. 1996, \apj, 463, 602
%
\reference{rutledge97} Rutledge, G.A., Hesser, J.E., Stetson, P.B.,
Mateo, M., Simard, L., Bolte, M., Friel, E.D., and Copin, Y. 1997, 
\pasp, submitted (Paper I)
%
\reference{Sandage67} Sandage, A., and Wildey, R. 1967, \apj, 150, 469
%
\reference{saranorris94} Sarajedini, A., and Norris, J.E. 1994, \apjs, 93, 161
%
\reference{smith84} Smith. H.A. 1984, \apj, 281, 148
%
\reference{sneden91} Sneden, C., Kraft, R.P., Prosser, C.F., and 
Langer, G.E. 1991, \aj, 102, 2001
%
\reference{sneden92} Sneden, C., Kraft, R.P., Prosser, C.F., and 
Langer, G.E. 1992, \aj, 104, 2121
%
\reference{sneden94} Sneden, C., Kraft, R.P., Langer, G.E., Prosser, C.F., and 
Shetrone, M.D. 1994, \aj, 107, 1773
%
\reference{spinrad69} Spinrad, H., and Taylor, B.J. 1969, \apj, 157, 1279
%
\reference{spinrad71} Spinrad, H., and Taylor, B.J. 1971, \apjs, 22, 445
%
\reference{stetson89ls} Stetson, P.B. 1989, Image and Data Processing, eds. B Barbuy,
E. Janot-Pacheco, A. M. Magalh\~aes, S.M. Viegas, published by Departamento Astron\^omia,
Instituto Astronomico e Geof\'isico, Universidade de S\~ao Paulo, C.P. 9638, S\~ao Paulo
01065, Brazil.
%
\reference{stetaas} Stetson, P.B., Hesser, J.E., McClure, R.D., van den
Bergh, S., Bolte, M., Harris, W.E., VandenBerg, D.A., Bond, H.E., Fahlman,
G.G., Richer, H.B., Bell, R.A. 1996, \baas, 28, 1362.
%
\reference{stetson96} Stetson, P.B., VandenBerg, D.A., and  Bolte, M. 1996, \pasp, 108, 560
%
\reference{suntzeff92} Suntzeff, N.B., Schommer, R.A., Olszewski,
E.W., Walker, A.R. 1992, \aj, 104 (5), 1743 (S92)
%
\reference{suntzeff93} Suntzeff, N.B., Mateo, M., Terndrup, D.M., Olszewski,
E.W., Geisler, D., and Weller, W. 1993, \apj, 418, 208 (S93)
%
\reference{suntzeff96} Suntzeff, N.B., and Kraft, R.P.  1996, 
\aj, 111, 1913 (SK96)
%
\reference{van65} van den Bergh, S., 1965, \jrasc, 59, 151
%
\reference{van67} van den Bergh, S., 1967, \aj, 72, 70
%
\reference{van93} van den Bergh, S., 1993, \aj, 105, 971
%
\reference{wheeler89} Wheeler, J.C., Sneden, C., and Truran, J.W. 1989, \araa, 27, 279
%
\reference{zinn80a} Zinn, R. 1980a, \apj, 241, 602
%
\reference{zinn80b} Zinn, R. 1980b, \apjs, 42, 19
%
\reference{zinn85} Zinn, R. 1985, \apj, 293, 424
%
\reference{zinn84} Zinn, R., and West, M.J. 1984, \apjs, 55, 45 (ZW84)

\end{references}
\end{document}